\documentclass{iaus}
\usepackage{graphicx}

\title[JD 11.~~] 
{The kinematical behavior of Galactic PNe with [WC] central star\footnote{Based on observations collected at OAN-SPM, Mexico and Las Campanas Observatory, Chile.}}
\author[ Rechy-Garc{\'\i}a,  Pe\~na \& Garc{\'\i}a-Rojas]   
{J.S Rechy-Garc{\'\i}a$^1$, Miriam Pe\~na$^2$
 \and Jorge Garc{\'\i}a-Rojas$^3$}
\affiliation{$^1$Fac. de Fisica e Inteligencia Artificial, U. Veracruzana, Xalapa, Ver., Mexico
 \\ email: {\tt jaci34@hotmail.com} \\[\affilskip]
$^2$Instituto de Astronomia, UNAM,   04510 Mexico D.F., email: {\tt miriam@astro.unam.mx\\}
$^3$ Instituto de Astrofisica de Canarias, La Laguna, Tenerife, Spain, email:{\tt jogarcia@iac.es}}

\pubyear{2011}
\volume{283}  
\pagerange{1--2}
\jname{Planetary Nebulae: An Eye to the Future}
\editors{A. Manchado, \& L. Stanghellini, eds.}
\begin{document}

\maketitle

\begin{abstract}
High resolution spectroscopic data of a large sample of galactic planetary nebulae with [WC] central stars ([WC]PNe) are analyzed to determine their kinematical behavior.
Their heliocentric velocities have been determined with a precision better than a few km/s. Distances obtained from the literature are used to derive the peculiar velocities of the objects.

\noindent Our preliminary results are: 
(a) The [WC]PNe are distributed in the galactic disk and they appear more concentrated than the normal PNe.  (b) Separating the sample in Peimbert's types, we find that Type I PNe show in general low peculiar velocities ($<$ 50 km/s) except for a couple of objects apparently belonging to the galactic bulge. For the other [WC]PNe, most of them belong to the Peimbert's Type II (defined as having  $V_{pec}$ $\leq$ 60 km/s). However there is an important fraction (28\%) showing  $V_{pec}$ larger than 60 km/s therefore they are classified as Peimbert's Type III. 
\keywords{ Galaxy: general, planetary nebulae: general, stars: AGB and post-AGB}
\end{abstract}

\firstsection 
\section{Introduction}

Galactic planetary nebulae (PNe)  belong mostly to the galactic disk. Within the central stars of PNe there is a particular group ($\approx$ 10\% of the known sample) which presents  W-R spectral type, i.e, they show large mass loss and an atmosphere of He, C and O ([WC] spectral type). We aim to study the kinematical behavior of this group. 

A sample of 72 PNe were observed of which 40 are [WC]PNe. The spectra were obtained with the spectrographs `echelle' of the OAN-SPM, M\'exico (47 objects) and MIKE at Clay 6.5-m telescope, Las Campanas Observatory (25 objects). The spectra cover a range between 3700 and 7000 \AA. To determine radial velocities IRAF\footnote{IRAF is distributed by the NOAO, which is operated by the AURA, Inc., under cooperative agreement  with the NSF.} tasks  were employed. 
We measured the central wavelengths of  the lines H$\gamma$, H$\beta$, H$\alpha$, [OIII]$\lambda\lambda$4959,5007 and [NII]$\lambda\lambda$6548, 6583 by performing a Gaussian fit. In the case of double profiles, we took the average of both components. The radial velocities were corrected for terrestrial motion, thus providing the heliocentric radial velocities. We also computed the circular velocitites, V$_{circ}$, corresponding to the heliocentric radial velocity due to galactic rotation. This allows us to determine the peculiar velocities of the objects, as given by:
\begin{equation}
V_{pec} = V_{helio} - V_{circ}
\end{equation}
To derived circular and peculiar velocities we used the distances by Stanghellini \& Haywood (2010).

\section{Galactic distribution}

Fig. 1  shows that most of the PNe are distributed in the galactic disk and they are strongly concentrated towards the galactic center. Many of them actually belong to the bulge.
[WC]PNe (Fig. 1 above) are more concentrated than normal PNe (Fig. 1 below), showing distances above the plane lower than 400 pc while normal PNe show typical distances  above the plane up to 800 pc. Most [WC]PN have a maximum  galactic latitude  b=$\pm 5^o$. In addition, a significant number  have been detected in the bulge (Gorny et al. 2009).
\begin{figure}[t] 
\begin{center}
\includegraphics[width=3.2in]{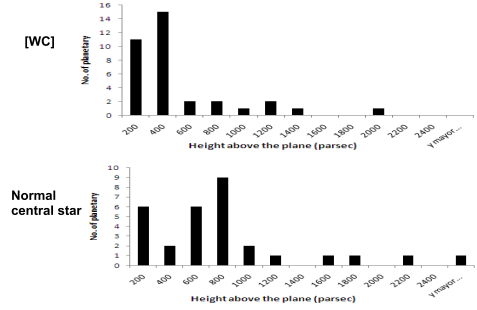}
\vskip -8pt
\caption{Distribution of distances relative to the galactic plane. Above: [WC]PNe, below: normal PNe.}
\end{center}
\end{figure}

\section{Conclusions}

- We have classified the [WC]PNe of our sample in Peimbert's types (Peimbert 1978). Of the 46  objects   analyzed, 8 are Peimbert's Type I (they are He and N-rich and represent 17.39\% of the total sample, two of them, PN G003.1+02.9 Hb4 and PN G011.9+04.2 M1-32, belong to the  bulge), 25 are Peimbert's Type II (defined as having  $V_{pec}$ $\leq$ 60 km/s, they represent the 54.35\%) and 13 are Peimbert's Type III (these have $V_{pec} > $ 60 km/s and represent the 28.26\%). Then most, being of Peimbert Type II, are middle age objects, and there are some [WC]PNe belonging to the old disk population.

- Two PNe (no [WC]) in our sample have very high radial velocity. One is the well known BoBn-1 (PN G108.4-76.1, V${pec}$=187.82 km/s) which belongs to the galactic halo, although recently it was found that it could belong to the Sagittarius dwarf galaxy (Zijlstra et al. 2006).  The other case, Hb12 (PN G111.8-02.8), shows an extremely large peculiar velocity (V${pec}=-$371.34 km/s) which classifies it as a very high velocity object despite it is located in the disk. For its high speed this object may belong to the halo.

\medskip This work was partially supported by DGAPA-UNAM, grant IN105511.


\begin{thebibliography}{}

\bibitem[]{}
Gorny, S. , Chiapini, C., Stasinska, G., Cuisinier, F., 2009, \textit{A\&A}, 500, 1089 
\bibitem[]{}
Peimbert, M., 1978, in:  Y. Terzian (ed.), \textit{Planetary Nebulae, Observations and Theory}, Proc. IAU Symp. No. 76 (Reidel Publishing Co.), p.\ 215
\bibitem[]{}
Stanghellini, L., \& Haywood, M., 2010, \textit{ApJ}, 714,  1096
\bibitem[]{}
Zijlstra, A. A.,  Gesicki, K., Walsh, R. J., et al., 2006, \textit{MNRAS}, 369, 875
\end{thebibliography}
\end{document}